\documentclass[conference,10pt]{IEEEtran}

\IEEEoverridecommandlockouts

\usepackage{cite}
\usepackage{amsmath,amssymb,amsfonts,amsthm}
\usepackage{mathtools}

\usepackage{graphicx}
\usepackage{xcolor}
\usepackage[normalem]{ulem}

\usepackage{pgf,tikz}
\usepackage{pgfplots}
\pgfplotsset{compat=newest}

\def\BibTeX{{\rm B\kern-.05em{\sc i\kern-.025em b}\kern-.08em
    T\kern-.1667em\lower.7ex\hbox{E}\kern-.125emX}}
    
\newtheorem{theorem}{Theorem}

\DeclareMathOperator{\tr}{tr}
\DeclareMathOperator{\mse}{mse}
\DeclareMathOperator{\mmse}{mmse}
\DeclareMathOperator{\MMSE}{MMSE}
\DeclareMathOperator{\SNR}{SNR}

\begin{document}

\title{Tight Bounds on the Weighted Sum of MMSEs\\ with Applications in Distributed Estimation
}

\author{\IEEEauthorblockN{Michael Fau{\ss}, Abdelhak M. Zoubir}
\IEEEauthorblockA{\textit{Signal Processing Group} \\
\textit{Technische Universit{\"a}t Darmstadt}\\
D-64283 Darmstadt, Germany \\
\{mfauss, zoubir\}@spg.tu-darmstadt.de}
\and
\IEEEauthorblockN{Alex Dytso, H. Vincent Poor}
\IEEEauthorblockA{\textit{Dept. of Electrical Engineering} \\
\textit{Princeton University}\\
Princeton, NJ 08544, USA \\
\{adytso, poor\}@princeton.edu}
\and
\IEEEauthorblockN{Nagananda Kyatsandra}
\IEEEauthorblockA{\textit{LTCI, T\'{e}l\'{e}com ParisTech} \\
\textit{Institut Mines - T\'{e}l\'{e}com}\\
Paris 75013, France \\
nkyatsandra@enst.fr}
}

\maketitle

\begin{abstract}
    In this paper, tight upper and lower bounds are derived on the weighted sum of minimum mean-squared errors for additive Gaussian noise channels. The bounds are obtained by constraining the input distribution to be close to a Gaussian reference distribution in terms of the Kullback--Leibler divergence. The distributions that attain these bounds are shown to be Gaussian whose covariance matrices are defined implicitly via systems of matrix equations. Furthermore, the estimators that attain the upper bound are shown to be minimax robust against deviations from the assumed input distribution. The lower bound provides a potentially tighter alternative to well-known inequalities such as the Cram\'{e}r--Rao lower bound. Numerical examples are provided to verify the theoretical findings of the paper. The results derived in this paper can be used to obtain performance bounds, robustness guarantees, and engineering guidelines for the design of local estimators for distributed estimation problems which commonly arise in wireless communication systems and sensor networks. 
\end{abstract}


\section{Introduction}

The mean square error (MSE) is a natural and commonly used measure for the accuracy of an estimator. The minimum MSE (MMSE) plays a central role in statistics \cite{Lehmann_1998, Dodge_2008}, information theory \cite{Guo_2011, Dytso_2018}, signal processing \cite{Kay_1993, Dalton_2012, Dalton_2012b}, and has close connections to entropy and mutual information \cite{Guo_2005, Guo_2006}. In \cite{FaussDytso2018_SSP} and \cite{FaussDytso2018_TSP}, lower and upper bounds on the MMSE are derived when the random variable of interest is contaminated by additive Gaussian noise and its distribution is $\varepsilon$-close to a Gaussian reference distribution in terms of the Kullback--Leibler (KL) divergence. The estimator that attains this upper bound is shown to be minimax robust in the sense that it minimizes the maximum MMSE over the set of feasible distributions. That is, within the specified  KL divergence ball, it is robust to arbitrary deviations of the prior from the nominal Gaussian case. The lower bound provides a fundamental limit on the estimation accuracy and is a potentially tighter alternative to the Bayesian Cram\'{e}r-Rao bound. 

This paper extends the bounds in \cite{FaussDytso2018_SSP} and \cite{FaussDytso2018_TSP} to a weighted sum of MMSEs. Similar to \cite{FaussDytso2018_SSP} and \cite{FaussDytso2018_TSP}, the bounds derived in this paper are obtained by constraining the input distribution to be $\varepsilon$-close to a Gaussian reference distribution in terms of the KL divergence. The estimators that attain the upper bound are minimax robust against deviations from the assumed input distribution.  Finally, the proposed  bounds are evaluated for generalized Gaussian distributions and the uniform distribution. It is shown that, in some cases, the lower bounds derived in this paper are tighter than the Cram\'{e}r--Rao lower bound. Interestingly, the performance of the proposed bounds improves as the dimension of the input vector increases.

The weighted sum of MMSEs arises in various practical applications in signal processing and communications. For example, it has been shown that for a Gaussian prior the MMSE in the time domain can be expressed as a sum of MMSEs in the frequency domain \cite{Avargel2007}. In multiple input multiple output (MIMO) wireless communications, the MMSE is frequently expressed as a sum of MMSEs or sum of inverse MMSEs of multiple parallel channels \cite{PrezCruz2007, Dietrich2008, Grundinger2012, Gonzalez-Coma2015}. In the context of distributed statistical inference, weighted sums of MMSEs play an important role in parameter estimation problems, where noisy measurements from multiple randomly deployed sensors are used to estimate the parameter of interest. In such scenarios, typically the estimates from the local sensors are fused to obtain a global estimate of the parameter. However, in practice, not much is known about the analytical characterization the optimal performance of the global estimator \cite{Zhang2016}. The study of the weighted sum of MMSEs reported in this paper is a step in this direction. The weighted sum of MMSEs is not only an appropriate objective function for distributed estimation, since it provides a platform to establish a global performance measure, but it also allows to prioritize sensors by assigning them weights. Thus, a highly informative sensor will be assigned a higher weight in the linear combination of MMSEs of all the sensors. This could have important applications in energy-efficient sensor networking where only highly informative sensors transmit their local decisions, while the sensors deemed less-informative abstain from transmission, thus saving energy and time for decision making \cite{Blum2011}, \cite{Sriranga2019}. 

The rest of the paper is organized as follows. In Section~\ref{sec:problem}, we provide a mathematical statement of the problem addressed in the paper. The upper and lower bounds on the weighted sum of MMSEs, which are the main results of this work, are stated in Section~\ref{sec:result}. The usefulness of the proposed bounds in distributed estimation is illustrated via an example application and related details are discussed in Section~\ref{sec:applications}. Concluding remarks are provided in Section~\ref{sec:conclusion}.


\section{Problem Formulation}
\label{sec:problem}
Let $(\mathbb{R}^K, \mathcal{B}^K)$ denote the $K$-dimensional Borel space. Consider $J$ additive-Gaussian-noise channels $Y_j = X + N_j$, $j = 1, \ldots, J$, where $X$ and $N_1, \ldots, N_J$ are independent $(\mathbb{R}^K, \mathcal{B}^K)$-valued random variables. All $N_j$ are assumed to be zero-mean Gaussian distributed, i.e., $P_{N_j} = \mathcal{N}(0, \Sigma_{N_j})$, where $\mathcal{N}(\mu,\Sigma)$ denotes the Gaussian distribution with mean $\mu$ and covariance $\Sigma$.

We define the individual MSEs as functions of the estimator $f_j$ and the input distribution $P_X$, i.e.,
\begin{equation*}
    \mse_{X | Y_j}(f_j, P_X) \coloneqq E_{P_{Y_j|X} P_X}\Bigl[ \left\lVert f_j(Y_j) - X \right\rVert^2\Bigr].
\end{equation*}
The individual MMSEs are accordingly defined as
\begin{equation*}
    \mmse_{X | Y_j}(P_X) \coloneqq \inf_{f_j \in \mathcal{F}} \; \mse_{X | Y_j}(f_j, P_X),
\end{equation*}
where $\mathcal{F}$ denotes the set of all all feasible estimators, i.e.,
\begin{equation*}
    \mathcal{F} = \left\{ f\colon (\mathbb{R}^K, \mathcal{B}^K) \to (\mathbb{R}^K, \mathcal{B}^K) \right\}.
\end{equation*}
The following two optimization problems are investigated in this paper:
\begin{align}
    \sup_{P_X \in \mathcal{P}_\varepsilon} \; \sum_{j=1}^J \lambda_j \mmse_{X|Y_j}(P_X) \quad &\text{s.t.} \quad P_X \in \mathcal{P}_\varepsilon, \label{eq:max_prob} \\
    \inf_{P_X \in \mathcal{P}_\varepsilon} \; \sum_{j=1}^J \lambda_j \mmse_{X|Y_j}(P_X) \quad &\text{s.t.} \quad P_X \in \mathcal{P}_\varepsilon, \label{eq:min_prob}
\end{align}
where $\lambda_1, \ldots, \lambda_J > 0$ are fixed positive weights and the set of feasible distribution is defined as 
\begin{equation}
    \label{eq:p_feasible}
    \mathcal{P}_\varepsilon \coloneqq \bigl\{ P_X : D_{\text{KL}}(P_X \,\Vert\, P_0) \leq \varepsilon \bigr\}.
\end{equation}
Note that\textcolor{orange}{\sout{,}} $\mathcal{P}_\varepsilon$ is a KL divergence ball centered at $P_0$ of radius $\varepsilon$. As we proceed, it will become clear that it is useful to choose the reference distribution $P_0$ to be Gaussian:
\begin{equation}
    \label{eq:P0}
    P_0 = \mathcal{N}(\mu_0,\Sigma_0).
\end{equation}
Finally, to allow for a compact notation, the following matrices are introduced:
\begin{align}
    W_j &\coloneqq \Sigma_N (\Sigma_X + \Sigma_{N_j})^{-1}, \\
    \MMSE_j &\coloneqq \Sigma_X \bigl( \Sigma_X + \Sigma_{N_j} \bigr)^{-1} \Sigma_{N_j} = \Sigma_X W^{\text{T}}, \label{eq:MMSE_matrix} \\
    \SNR_0 &\coloneqq \Sigma_0^{-1} \Sigma_X,
\end{align}
where $j = 1,\ldots,J$ and $W^\text{T}$ denotes the transpose of $W$.

\section{Main Result}
\label{sec:result}
The main result of the paper is provided in the following theorem.
\begin{theorem}
    If $(\alpha^* ,\Sigma_X^*)$, with $\Sigma_X^*$ positive definite and $\alpha^* \geq 0$, solve
    \begin{gather}
      \label{eq:opt_covariance}
      \Sigma_X = \Sigma_0 + \alpha \left( \sum_{j=1}^J \lambda_j \MMSE_j^{\text{T}} \MMSE_j \right) \SNR_0^{-1} \\[1ex]
      \label{eq:opt_KL}
      \tr\bigl( \SNR_0 \bigr) - \log\det\bigl( \SNR_0 \bigr) - K = 2\varepsilon,
    \end{gather}
    then 
    \begin{equation}
      \label{eq:px_opt_sup}
      P_X^* = \mathcal{N}\bigl(\mu_0, \Sigma_X^* \bigr)
    \end{equation}
    solves problem \eqref{eq:max_prob}. Analogously, if $(\alpha^\dagger ,\Sigma_X^\dagger)$, with $\Sigma_X^\dagger$ positive definite and $\alpha^\dagger \leq 0$, solve \eqref{eq:opt_covariance} and \eqref{eq:opt_KL}, then 
    \begin{equation}
      \label{eq:px_opt_inf}
      P_X^\dagger = \mathcal{N}\bigl(\mu_0, \Sigma_X^\dagger \bigr)
    \end{equation}
    solves problem \eqref{eq:min_prob}.
\end{theorem}

Since the input distributions that attain these bounds are Gaussian of the form $P_X \sim \mathcal{N}\bigl(\mu_0, \Sigma_X \bigr)$, the individual MMSE estimators in both cases are given by
\begin{equation}
    \label{eq:MMSE_estimator}
    f_j(y_j) = (I-W_j) y_j + W_j \mu_0,
\end{equation}
where $y_j \in \mathbb{R}^K$ denotes the observed realization of $Y_j$. The corresponding MMSEs are given by
\begin{align}
    \label{eq:MMSE}
    \mmse_{X|Y_j}(P_X) &= \tr\Bigl( \Sigma_X (\Sigma_X + \Sigma_{N_j})^{-1} \Sigma_{N_j} \Bigr) \\
  &= \tr\bigl( \MMSE_j \bigr).
\end{align}
The lower and upper bounds on the weighted sum of MMSEs in \eqref{eq:max_prob} and \eqref{eq:min_prob} are then given by
\begin{align}
    \label{eq:MMSE_upper_bound}
    \sum_{j=1}^J \lambda_j \mmse_{X|Y_j}(P_X) \leq \sum_{j=1}^J \lambda_j \tr\Bigl( \MMSE_j^* \Bigr)
    \intertext{and}
    \label{eq:MMSE_lower_bound}
    \sum_{j=1}^J \lambda_j \mmse_{X|Y_j}(P_X) \geq \sum_{j=1}^J \lambda_j \tr\Bigl( \MMSE_j^\dagger \Bigr),
\end{align}
where $\MMSE_j^*$ and $\MMSE_j^\dagger$ are shorthand notations for $\MMSE_j$ in \eqref{eq:MMSE_matrix} evaluated at $\Sigma_X^*$ and $\Sigma_X^\dagger$, respectively.

\begin{proof}
The proof of the main result follows along the same lines as the proof in \cite[Sec.~4]{FaussDytso2018_SSP}. Consider the maximization in \eqref{eq:max_prob}, which can be written as the minimax problem
\begin{equation}
    \label{eq:nested_problem}
    \sup_{P_X \in \mathcal{P}_\varepsilon} \; \inf_{f_j \in \mathcal{F}} \; \sum_{j=1}^J \lambda_j \mse_{X|Y_j}(f_j,P_X),
\end{equation}
where the infimum is taken jointly over all $f_1, \ldots, f_J$. A sufficient condition for $P_X^*$ and $f_1^*, \ldots, f_J^*$ to solve \eqref{eq:nested_problem}, and hence \eqref{eq:max_prob}, is that they satisfy the saddle point conditions \cite[Exercise 3.14]{Boyd_2004}
\begin{equation}
    \label{eq:saddle_point_f} 
    \sum_{j=1}^J \lambda_j \mse_{X|Y_j}(f_j^*,P_X^*) \leq \sum_{j=1}^J \lambda_j \mse_{X|Y_j}(f_j,P_X^*)
\end{equation}
for all $f_1, \ldots, f_J \in \mathcal{F}$ and
\begin{equation}
    \label{eq:saddle_point_P} 
    \sum_{j=1}^J \lambda_j \mse_{X|Y_j}(f_j^*,P_X^*) \geq \sum_{j=1}^J \lambda_j \mse_{X|Y_j}(f_j^*,P_X)
\end{equation}
for all $P_X \in \mathcal{P}_\varepsilon$. The fact that the estimators $f_j$ in \eqref{eq:MMSE_estimator} minimize the right hand side of \eqref{eq:saddle_point_f} follows directly from the definition of the MMSE \cite[Chapter 10.4]{Penny_2000}. In the remainder of the proof, it is shown that $P_X^*$ in \eqref{eq:px_opt_sup} satisfies \eqref{eq:saddle_point_P}.  

First, the right hand side of \eqref{eq:saddle_point_P} is written as
\begin{equation*}
    \sum_{j=1}^J \lambda_j \mse_{X|Y_j}(f_j^*,P_X) = E_{P_X} \bigl[ h(X) \bigr],
\end{equation*}
where $h\colon \mathbb{R}^K \to \mathbb{R}$ is independent of $P_X$ and given by
\begin{align}
    h(x) &= \sum_{j=1}^J \lambda_j E_{P_{Y_j \mid X=x}} \Bigl[ \left\lVert f_j^*(Y_j) - x \right\rVert_2^2 \Bigr] \\
    &= \sum_{j=1}^J \lambda_j E_{\mathcal{N}(x, \Sigma_{N_j})} \Bigl[ \left\lVert (I-W_j) Y_j + W_j \mu_0 - x \right\rVert_2^2 \Bigr] \\
    &= c + \sum_{j=1}^J \lambda_j \left\lVert (I-W_j) x + W_j \mu_0 - x \right\rVert_2^2 \\
    &= c + \sum_{j=1}^J \lambda_j (x - \mu_0)^\text{T} W_j^\text{T} W_j (x-\mu_0) \\
    &= c + (x - \mu_0)^\text{T} \left(\sum_{j=1}^J \lambda_j  W_j^\text{T} W_j \right) (x-\mu_0),
    \label{eq:h_func}
\end{align}
with $c \coloneqq \sum_{j=1}^J \lambda_j \tr \bigl( (I-W_j) \Sigma_N (I-W_j)^\text{T} \bigr)$
being a constant that is independent of $x$. Using the auxiliary result on bounds on expectations under $f$-divergence constraints in \cite[Sec.~4.1]{FaussDytso2018_SSP}, it follows that the density $p_X^*$ of the optimal distribution $P_X^*$ is of the form
\begin{equation}
    \label{eq:px_opt}
    p_X^*(x) = p_0(x) \; e^{\alpha h(x) + \beta - 1},
\end{equation}
where $p_0$ denotes the density of the reference distribution $P_0$ and $\alpha \geq 0$, $\beta \in \mathbb{R}$ need to be chose appropriately. Substituting $h$ in \eqref{eq:px_opt} with \eqref{eq:h_func} and using \eqref{eq:P0} yields
\begin{align}
    p_X^*(x) &\propto p_0(x) \; e^{\alpha (x - \mu_0)^\text{T} \bigl( \sum_{j=1}^J \lambda_j  W_j^\text{T} W_j \bigr) (x-\mu_0) } \\ 
    &\propto e^{ -\frac{1}{2} (x - \mu_0)^\text{T} \bigl( \Sigma_0^{-1} - \alpha \sum_{j=1}^J \lambda_j  W_j^\text{T} W_j \bigr) (x-\mu_0) } \\
    &= e^{ -\frac{1}{2} (x - \mu_0)^\text{T} \Sigma_X^{-1}  (x-\mu_0) },
\end{align}
where, without loss of generality, $\alpha$ has been scaled by $\tfrac{1}{2}$ and 
\begin{equation}
    \label{eq:opt_condition}
    \Sigma_X^{-1} = \Sigma_0^{-1} - \alpha \sum_{j=1}^J \lambda_j  W_j^\text{T} W_j.
\end{equation}
Multiplying both sides of \eqref{eq:opt_condition} by $\Sigma_X$ from the left and the right and rearranging the terms yields \eqref{eq:opt_covariance}.

Knowing that $P_X$ and $P_0$ are Gaussian distributions with identical means, their KL divergence is given by 
\begin{equation}
    \label{eq:gauss_kl}
    D_{\text{KL}}(P_X \,\Vert\, P_0) = \frac{1}{2}\left( \tr\bigl( \SNR_0 \bigr) - K - \log\det\bigl( \SNR_0 \bigr) \right).
\end{equation}
Equating \eqref{eq:gauss_kl} with $\varepsilon$ yields the optimality condition \eqref{eq:opt_KL}. This concludes the proof of optimality of $P_X^*$.

The proof of optimality of $P_X^\dagger$ follows analogously, the only difference being that the sign of $\alpha$ is reversed; relevant details are provided in  \cite[Sec.~4.1]{FaussDytso2018_SSP}.
\end{proof}

\section{An application involving distributed estimation} 
\label{sec:applications}

In this section, we first present an example application involving distributed estimation where the weighted sum of MMSEs is relevant. We show how a simple modification of the conventional distributed processing leads to significant improvements in the performance characterization of practical distributed estimators. We then derive bounds on the weighted sum of MMSEs for distributed estimation with arbitrary distributions in Gaussian noise. Lastly, we specialize these bounds for the generalized Gaussian and uniform distributions. Numerical evaluations, presented to verify the theoretical findings, reveal surprising features of the proposed bounds. 

Consider the problem of distributed estimation using wireless sensor networks (WSNs), where $J$ sensors are randomly distributed in the region of interest (ROI). The $j^{\text{th}}$ sensor is located at a distance $d_j$ from a target. By ``target'' we are referring to some activity; for example, fire in the ROI. The target's signal power is assumed to follow the isotropic power attenuation model \cite{Sriranga2018}. The signal power at sensor $j$ is given by $\rho_j^2 = \rho_0^2 /(1 + \gamma d_j^m)$, $j = 1, \dots, J$, where $\rho_0^2$ is the target's signal power at distance zero, $m$ is the signal decay exponent taking values between 2 and 3, and $\gamma$ is a constant (larger $\gamma$ implies faster power decay). 
In practice, the parameters $m$ and $\gamma$ pertaining to the wireless medium are obtained by performing experiments before the WSN is deployed, though uncertainty is associated with this knowledge. For the purpose of this paper, let us consider the simple goal of estimating the distances $d_j$ based on the noisy observations made by the sensors. The knowledge of $d_j$ is typically used to infer the presence/absence of a target in the ROI (see \cite{Sriranga2018}). In conventional distributed estimation, the estimates of $d_j$ computed by the $j^{\text{th}}$ local sensor is  transmitted to a central processing unit, which aggregates $d_j$, $j = 1, \dots, J$ to compute a system-level estimate of the distance vector $\hat{\mathbf{d}} = (\hat{d}_1, \dots, \hat{d}_J)$. However, to the best of our knowledge, theoretical insights into the system-level estimator's accuracy are lacking in the literature.

Let us now consider a simple modification to the above scheme. Instead of the local estimates, if the $J$ sensors transmit their local MSEs to the central unit, then a weighted sum of MMSEs can be thoroughly analyzed using the findings of this paper. This new scheme provides a comprehensive view of the performance of system-level (or, global) estimators unlike existing distributed estimation wherein the local estimates are simply fused at the control unit without insightful performance guarantees. Our study provides engineering guidelines for the design and analysis of large-scale sensor networks which are important components in several critical infrastructures like the Smart Grid, IoT and other cyber-physical systems. Possible improvements in global system performance are demonstrated with the following two examples. We first derive upper and lower bounds on the linear combination of MMSEs for arbitrary distributions in Gaussian noise and then derive these bounds for the generalized Gaussian distribution. 

\subsection{Bounds on the linear combinations of MMSEs for arbitrary distributions in Gaussian noise} 
Consider the linear combination of MMSEs for an arbitrary distribution $P_X$ such that 
\begin{align}
 \min_Q D_{\text{KL}}(P_X\|Q)<\infty. 
\end{align}

For a given $P_X$, upper and lower bounds on the linear combination of MMSEs can be derived by the following steps:
\begin{enumerate}
    \item Find the best Gaussian approximation of $P_X$ in terms of the KL divergence. In other worlds, find a Gaussian $Q$ that minimizes $D_{\text{KL}}(P_X\|Q)$ and compute $\varepsilon=D_{\text{KL}}(P_X\|Q) $.
    \item  Use the value of $\varepsilon$ found in Step 1 to compute the upper and lower bounds in \eqref{eq:max_prob} and \eqref{eq:min_prob}, respectively. 
\end{enumerate}

We can evaluate the effectiveness of this procedure by comparing it to the bounds attained by individually bounding each term in the linear combination. The most popular bounds on the individual MMSEs are the following:
\begin{align}
    \mmse_{X|Y_j}(P_X) &\le \tr \left( \Sigma_X (\Sigma_X+  \Sigma_{N_j} )^{-1} \Sigma_{N_j} \right) , \label{eq:LMMSE}\\
    \mmse_{X|Y_j}(P_X)  &\ge \frac{K^2}{\tr \left(\Sigma_{N_j}^{-1} \right)+ J(P_X) }, \label{eq:CRlower}
\end{align}
where $J(P_X)$ is the Fisher information of $P_X$. 
The upper bound in \eqref{eq:LMMSE} is obtained by  using the best linear estimator instead of the optimal estimator. The lower bound in \eqref{eq:CRlower} is the  Cram\'er--Rao lower bound. We refer to the bounds obtained by bounding individual MMSEs as local and the bounds that work directly on the linear combination as global. 

\subsection{Generalized Gaussian and Uniform distributions}
Let us now consider distributions that are either concentrated or heavy-tailed. A classic example is the generalized Gaussian distribution, whose density is given by 
\begin{align}
    f_X(x)=  c_p  {\rm e}^{-\frac{\| x\|^p}{p}},
\end{align}
where $c_p$ is the normalization constant. The covariance matrix, Fisher information and the best Gaussian approximation for this distribution are given by
\begin{align}
    \Sigma_X &= \frac{p^\frac{2}{p} \Gamma \left(  \frac{K+2}{p} \right)}{K \Gamma \left(  \frac{K}{p} \right)}  I ,\\
    J(X) &= p^{\frac{2p-2}{p}} \frac{\Gamma \left(  \frac{K+2p-2}{p} \right)}{\Gamma \left(  \frac{K}{p} \right)},\\
    \varepsilon_p &= \min_Q D_{\text{KL}}(P_X\|Q)\\
    &= \frac{n}{2}-\frac{n}{p}-\frac{n}{p}\log(p) - \log\left( \frac{ \pi^{\frac{n}{2}} \Gamma \left( \frac{n}{p}+1 \right) }{\Gamma \left( \frac{n}{2}+1 \right)} \right) \notag \\
    & \quad + \frac{n}{2} \log \left( \frac{2 \pi p^{\frac{2}{p}} \Gamma \left( \frac{n+2}{p} \right)}{n \Gamma \left( \frac{n}{p} \right) }\right).
\end{align}

\begin{figure}[t]
  \centering
%
%
\definecolor{mycolor1}{rgb}{0.50196,0.50196,0.50196}%
\begin{tikzpicture}

\begin{axis}[%
  width=225pt,
    height=175pt,
    at={(1.011in,0.642in)},
scale only axis,
xmin=0.0403225806451615,
xmax=10,
xlabel style={font=\color{white!15!black}},
 xlabel={$p$},,
ymode=log,
ymin=0.465460907249561,
ymax=21.4840813573167,
yminorticks=true,
axis background/.style={fill=white},
xmajorgrids,
ymajorgrids,
yminorgrids,
legend style={legend cell align=left, align=left, draw=white!15!black}
]
\addplot [color=black, line width=1.0pt]
  table[row sep=crcr]{%
0.51	16.5573685379054\\
0.905416666666667	8.38118348765315\\
1.30083333333333	5.25453080819775\\
1.69625	3.97986526437939\\
2.09166666666667	3.25619447033735\\
2.48708333333333	2.63453059314903\\
2.8825	2.23555444799002\\
3.27791666666667	1.96080770495218\\
3.67333333333333	1.76131096812669\\
4.06875	1.61040938353516\\
4.46416666666667	1.49252172361581\\
4.85958333333333	1.39799859857884\\
5.255	1.32057490748002\\
5.65041666666667	1.25601965274366\\
6.04583333333333	1.20137990122485\\
6.44125	1.15453522212419\\
6.83666666666667	1.11392616372372\\
7.23208333333333	1.0783811665292\\
7.6275	1.0470037307682\\
8.02291666666667	1.01909656419565\\
8.41833333333333	0.994109401415954\\
8.81375	0.971602338497281\\
9.20916666666667	0.951219575850372\\
9.60458333333333	0.932670290529425\\
10	0.915714484878532\\
};
\addlegendentry{Proposed Lower Bound}

\addplot [color=black, dotted, line width=1.0pt]
  table[row sep=crcr]{%
0.51	19.5414711726452\\
0.905416666666667	11.10311382412\\
1.30083333333333	6.40442876643134\\
1.69625	4.32221430172409\\
2.09166666666667	3.33240003488259\\
2.48708333333333	2.95080905951131\\
2.8825	2.70230937569996\\
3.27791666666667	2.52870391404924\\
3.67333333333333	2.40098012607036\\
4.06875	2.30322060443261\\
4.46416666666667	2.22602751908604\\
4.85958333333333	2.16352201493395\\
5.255	2.1118512804888\\
5.65041666666667	2.06839147389728\\
6.04583333333333	2.03129745038108\\
6.44125	1.99923598246926\\
6.83666666666667	1.97122026788162\\
7.23208333333333	1.94650573914235\\
7.6275	1.92451994811564\\
8.02291666666667	1.90481588751901\\
8.41833333333333	1.88703955217994\\
8.81375	1.87090702542589\\
9.20916666666667	1.85618799678194\\
9.60458333333333	1.84269369682989\\
10	1.83026794508421\\
};
\addlegendentry{Proposed Upper Bound}

\addplot [color=mycolor1, line width=1.0pt]
  table[row sep=crcr]{%
0.51	18.920021648083\\
0.905416666666667	9.9453526063935\\
1.30083333333333	5.85143124888336\\
1.69625	4.15214201669418\\
2.09166666666667	3.29431633789706\\
2.48708333333333	2.79247145452451\\
2.8825	2.46751336400264\\
3.27791666666667	2.24139592990721\\
3.67333333333333	2.07549303363589\\
4.06875	1.94875482781112\\
4.46416666666667	1.84882285976554\\
4.85958333333333	1.76800079070013\\
5.255	1.70126160882243\\
5.65041666666667	1.64519029965522\\
6.04583333333333	1.59738940085391\\
6.44125	1.55612819838558\\
6.83666666666667	1.52012701386854\\
7.23208333333333	1.48841979473254\\
7.6275	1.46026386199633\\
8.02291666666667	1.4350790213687\\
8.41833333333333	1.41240550124452\\
8.81375	1.3918742790454\\
9.20916666666667	1.37318574985872\\
9.60458333333333	1.35609413043819\\
10	1.34039588069404\\
};
\addlegendentry{Local Upper Bound via \eqref{eq:LMMSE}}

\addplot [color=mycolor1, dashed, line width=1.0pt]
  table[row sep=crcr]{%
0.51	3.89466630455799\\
0.905416666666667	3.10585074173134\\
1.30083333333333	2.53553220621304\\
1.69625	2.14127316096058\\
2.09166666666667	1.85486966602855\\
2.48708333333333	1.63703443751596\\
2.8825	1.46539242988021\\
3.27791666666667	1.32644236275312\\
3.67333333333333	1.21155147169549\\
4.06875	1.11492398716895\\
4.46416666666667	1.03251077085182\\
4.85958333333333	0.961389405313789\\
5.255	0.899393224760732\\
5.65041666666667	0.84487932578576\\
6.04583333333333	0.796577941251732\\
6.44125	0.753491454874543\\
6.83666666666667	0.714824805490622\\
7.23208333333333	0.679936366390009\\
7.6275	0.648302546975773\\
8.02291666666667	0.619491812213767\\
8.41833333333333	0.593145302597302\\
8.81375	0.568962167091198\\
9.20916666666667	0.546688317772722\\
9.60458333333333	0.526107706120577\\
10	0.507035482949516\\
};
\addlegendentry{Local Lower Bound via \eqref{eq:CRlower} }

\addplot [color=black, dashdotted, line width=1.0pt]
  table[row sep=crcr]{%
0.51	13.1163874883813\\
0.905416666666667	6.99624157817574\\
1.30083333333333	4.80798913637855\\
1.69625	3.86039914837469\\
2.09166666666667	3.23158817416725\\
2.48708333333333	2.5446522552591\\
2.8825	2.11530572226592\\
3.27791666666667	1.8254242551417\\
3.67333333333333	1.61813315685518\\
4.06875	1.46324354582344\\
4.46416666666667	1.34344947067871\\
4.85958333333333	1.24820126713846\\
5.255	1.17073852436594\\
5.65041666666667	1.10654646820362\\
6.04583333333333	1.0525042400448\\
6.44125	1.00638958484745\\
6.83666666666667	0.966580264054401\\
7.23208333333333	0.931865557885781\\
7.6275	0.901324453074899\\
8.02291666666667	0.87424438537129\\
8.41833333333333	0.850065731205346\\
8.81375	0.828343052070241\\
9.20916666666667	0.808717501760293\\
9.60458333333333	0.79089683282505\\
10	0.774640677215008\\
};
\addlegendentry{Proposed Lower Bound Local Version}

\addplot [color=black, dashed, line width=1.0pt]
  table[row sep=crcr]{%
0.51	20.3146738593811\\
0.905416666666667	12.5899088523282\\
1.30083333333333	6.94803734154963\\
1.69625	4.45316866917787\\
2.09166666666667	3.35765343193734\\
2.48708333333333	3.05274287886461\\
2.8825	2.85033598218022\\
3.27791666666667	2.70704771513949\\
3.67333333333333	2.60062601741582\\
4.06875	2.51859563982409\\
4.46416666666667	2.45346798770221\\
4.85958333333333	2.40049701509416\\
5.255	2.35654193438697\\
5.65041666666667	2.31944673453566\\
6.04583333333333	2.28768599298565\\
6.44125	2.26015211034806\\
6.83666666666667	2.23602267052436\\
7.23208333333333	2.2146749711141\\
7.6275	2.19562934534175\\
8.02291666666667	2.17851061847476\\
8.41833333333333	2.16302130716867\\
8.81375	2.14892261040574\\
9.20916666666667	2.1360206892211\\
9.60458333333333	2.12415659478504\\
10	2.11319877750375\\
};
\addlegendentry{Proposed Upper Bound Local Version}

\end{axis}
\end{tikzpicture}%
  \caption{ Comparing local bounds to global bounds   }
  \label{fig:GlobalBoundsVs.Local}
\end{figure}
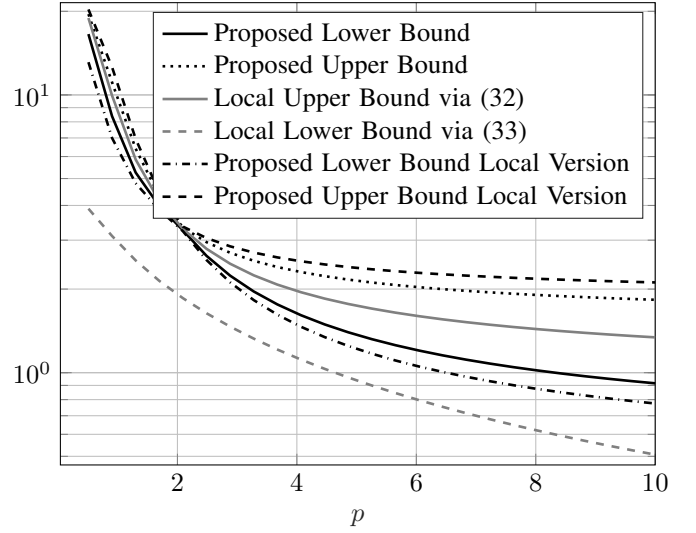

To evaluate the performance of our bounds we set 
 $ K = 3$, $J=4$,   and 
 \begin{subequations}
\begin{align}
\lambda&=\begin{pmatrix*}[r] 0.3565 & 0.0732 & 0.5910 & 0.9102 \end{pmatrix*}\\
    \Sigma_{N_1}&= \begin{pmatrix*}[r]
      \; 3.0405 &  -2.1179 &   2.1107 \;\\
 \;  -2.1179  &  4.1238  & -1.3414 \;\\
  \;  2.1107 &  -1.3414  &  4.8199 \;\\
        \end{pmatrix*}  \\
    \Sigma_{N_2}&= \begin{pmatrix*}[r]
     \; 0.9221 &   1.2047  &  0.5731 \;\\
  \;  1.2047  &  2.3851 &  -0.2188 \;\\
    \; 0.5731 &  -0.2188 &   1.5767 \;\\
        \end{pmatrix*}  \\
    \Sigma_{N_3}&= \begin{pmatrix*}[r]
     \;  9.9708  &  0.7749 &  -2.4323\;\\
    \;  0.7749  &  0.9252 &  -2.3907\;\\
   \;  -2.4323 &  -2.3907 &   6.3022\;\\
        \end{pmatrix*} \\
          \Sigma_{N_4}&= \begin{pmatrix*}[r]
   \;  1.2353 &  -1.1973 &  -1.1141 \;\\
 \;  -1.1973  &  4.2225  &  1.0695 \;\\
 \;  -1.1141  &  1.0695 &   1.6102 \;\\
        \end{pmatrix*} 
\end{align}
\label{eq:channelParameters}
\end{subequations}
and compare the resulting bounds in Fig.~\ref{fig:GlobalBoundsVs.Local}. Specifically, Fig.~\ref{fig:GlobalBoundsVs.Local} comprises the following bounds:
\begin{enumerate}
    \item lower bounds obtained via \eqref{eq:min_prob} (solid black line);
    \item upper bounds obtained via \eqref{eq:max_prob} (dotted black line); 
    \item   local upper bounds attained via \eqref{eq:LMMSE} (solid gray line);
    \item   local lower bounds attained via  \eqref{eq:CRlower} (dashed gray line);
    \item local lower bounds attained via \eqref{eq:min_prob} (dashed-dotted black line). The local version of the lower bound in \eqref{eq:min_prob} is obtained by individually minimizing each MMSE with the same KL constraint. Hence, the resulting solution is independent of $\lambda_i$'s; and
    \item local upper bounds attained via \eqref{eq:max_prob} (dashed black line). The local version of the upper bound in \eqref{eq:max_prob} is obtained by individually maximizing each MMSE with the same KL constraint. 
\end{enumerate}

\begin{figure}[t]
  \centering
%
%
\begin{tikzpicture}

\begin{axis}[%
width=220pt,
    height=175pt,
    at={(1.011in,0.642in)},
scale only axis,
xmin=0,
xmax=40,
xlabel style={font=\color{white!15!black}},
xlabel={$R$},
ymin=0,
ymax=25,
ylabel style={font=\color{white!15!black}},
axis background/.style={fill=white},
xmajorgrids,
ymajorgrids,
legend style={legend cell align=left, align=left, draw=white!15!black,at={(0.3,0.5)},anchor=west}
]
\addplot [color=black, line width=1.0pt]
  table[row sep=crcr]{%
0.1	0.0632015949047764\\
0.605063291139241	1.75407455293612\\
1.11012658227848	4.52307249033371\\
1.61518987341772	7.37830960400762\\
2.12025316455696	9.90337797934136\\
2.6253164556962	11.9892861756702\\
3.13037974683544	13.6584043615912\\
3.63544303797468	14.9768773798145\\
4.14050632911392	16.0162302744636\\
4.64556962025316	16.838947024174\\
5.1506329113924	17.4950650705609\\
5.65569620253164	18.0230961190221\\
6.16075949367089	18.4521824492532\\
6.66582278481013	18.8042680065\\
7.17088607594937	19.0959048711044\\
7.67594936708861	19.3396458872706\\
8.18101265822785	19.5450818354822\\
8.68607594936709	19.719602402298\\
9.19113924050633	19.868951892391\\
9.69620253164557	19.9976354890515\\
10.2012658227848	20.1092175618183\\
10.7063291139241	20.2065420911972\\
11.2113924050633	20.2918967469806\\
11.7164556962025	20.3671359919226\\
12.2215189873418	20.4337741949546\\
12.726582278481	20.4930566337258\\
13.2316455696203	20.5460140708862\\
13.7367088607595	20.593505031837\\
14.2417721518987	20.6362488026938\\
14.746835443038	20.6748513725468\\
15.2518987341772	20.7098259709204\\
15.7569620253165	20.7416094350169\\
16.2620253164557	20.7705753367819\\
16.7670886075949	20.7970445753965\\
17.2721518987342	20.8212939744037\\
17.7772151898734	20.8435632979845\\
18.2822784810127	20.8640610075673\\
18.7873417721519	20.8829690087732\\
19.2924050632911	20.9004465847435\\
19.7974683544304	20.9166336703631\\
20.3025316455696	20.9316535898719\\
20.8075949367089	20.9456153554857\\
21.3126582278481	20.958615605232\\
21.8177215189873	20.9707402429582\\
22.3227848101266	20.9820658314432\\
22.8278481012658	20.992660779994\\
23.3329113924051	21.0025863603074\\
23.8379746835443	21.0118975782812\\
24.3430379746835	21.0206439245612\\
24.8481012658228	21.0288700226495\\
25.353164556962	21.0366161901877\\
25.8582278481013	21.0439189264105\\
26.3632911392405	21.0508113366262\\
26.8683544303797	21.0573235028223\\
27.373417721519	21.0634828080464\\
27.8784810126582	21.069314221017\\
28.3835443037975	21.0748405464237\\
28.8886075949367	21.0800826455509\\
29.3936708860759	21.0850596311681\\
29.8987341772152	21.0897890400523\\
30.4037974683544	21.094286986021\\
30.9088607594937	21.0985682959465\\
31.4139240506329	21.1026466308756\\
31.9189873417722	21.1065345940849\\
32.4240506329114	21.1102438276556\\
32.9291139240506	21.1137850989366\\
33.4341772151899	21.1171683780866\\
33.9392405063291	21.1204029077291\\
34.4443037974684	21.1234972656227\\
34.9493670886076	21.1264594211335\\
35.4544303797468	21.1292967862003\\
35.9594936708861	21.1320162613953\\
36.4645569620253	21.134624277614\\
36.9696202531646	21.137126833857\\
37.4746835443038	21.1395295315215\\
37.979746835443	21.1418376055602\\
38.4848101265823	21.1440559528344\\
38.9898734177215	21.1461891579433\\
39.4949367088608	21.1482415167832\\
40	21.1502170580621\\
};
\addlegendentry{Proposed Upper Bound}

\addplot [color=black, dashed, line width=1.0pt]
  table[row sep=crcr]{%
0.1	0.0146361957443295\\
0.605063291139241	0.455081736173895\\
1.11012658227848	1.34081103063086\\
1.61518987341772	2.48243342510919\\
2.12025316455696	3.74727870263932\\
2.6253164556962	5.04449433960301\\
3.13037974683544	6.31712936343056\\
3.63544303797468	7.53070487824875\\
4.14050632911392	8.66509239905846\\
4.64556962025316	9.70993251375262\\
5.1506329113924	10.6618988560987\\
5.65569620253164	11.5228160045509\\
6.16075949367089	12.2980723944499\\
6.66582278481013	12.9950606549486\\
7.17088607594937	13.6218075576006\\
7.67594936708861	14.1860791857184\\
8.18101265822785	14.6949742858902\\
8.68607594936709	15.1548191909683\\
9.19113924050633	15.5711947919693\\
9.69620253164557	15.9490059949963\\
10.2012658227848	16.2925579871943\\
10.7063291139241	16.6056279139918\\
11.2113924050633	16.8915294624649\\
11.7164556962025	17.1531702825574\\
12.2215189873418	17.3931029714144\\
12.726582278481	17.6135702300279\\
13.2316455696203	17.8165447445456\\
13.7367088607595	18.0037642623741\\
14.2417721518987	18.1767622930645\\
14.746835443038	18.3368948480118\\
15.2518987341772	18.4853636262345\\
15.7569620253165	18.6232360451796\\
16.2620253164557	18.751462500216\\
16.7670886075949	18.870891213079\\
17.2721518987342	18.9822809994907\\
17.7772151898734	19.0863122521027\\
18.2822784810127	19.1835963994762\\
18.7873417721519	19.2746840672307\\
19.2924050632911	19.3600721352621\\
19.7974683544304	19.4402098559159\\
20.3025316455696	19.5155041725366\\
20.8075949367089	19.5863243558888\\
21.3126582278481	19.65300605731\\
21.8177215189873	19.7158548617754\\
22.3227848101266	19.7751494109297\\
22.8278481012658	19.8311441552068\\
23.3329113924051	19.8840717850468\\
23.8379746835443	19.9341453836445\\
24.3430379746835	19.9815603373423\\
24.8481012658228	20.0264960345046\\
25.353164556962	20.069117379294\\
25.8582278481013	20.1095761430537\\
26.3632911392405	20.1480121728765\\
26.8683544303797	20.184554474291\\
27.373417721519	20.2193221827513\\
27.8784810126582	20.2524254367041\\
28.3835443037975	20.2839661633713\\
28.8886075949367	20.3140387869839\\
29.3936708860759	20.3427308680006\\
29.8987341772152	20.3701236808003\\
30.4037974683544	20.3962927364418\\
30.9088607594937	20.4213082563019\\
31.4139240506329	20.4452356017249\\
31.9189873417722	20.4681356642258\\
32.4240506329114	20.490065220273\\
32.9291139240506	20.5110772542232\\
33.4341772151899	20.5312212525869\\
33.9392405063291	20.5505434724521\\
34.4443037974684	20.5690871865895\\
34.9493670886076	20.5868929074887\\
35.4544303797468	20.6039985923428\\
35.9594936708861	20.6204398307787\\
36.4645569620253	20.6362500169506\\
36.9696202531646	20.6514605074455\\
37.4746835443038	20.6661007663017\\
37.979746835443	20.6801984983107\\
38.4848101265823	20.6937797716577\\
38.9898734177215	20.7068691308485\\
39.4949367088608	20.7194897007812\\
40	20.7316632827372\\
};
\addlegendentry{Proposed Lower Bound}

\addplot [color=black, dotted, line width=1.0pt]
  table[row sep=crcr]{%
0.1	0.0333385268707848\\
0.605063291139241	0.983623635397363\\
1.11012658227848	2.73248889597286\\
1.61518987341772	4.78224589952632\\
2.12025316455696	6.8339948360954\\
2.6253164556962	8.73231894847925\\
3.13037974683544	10.4133632055481\\
3.63544303797468	11.8652137703321\\
4.14050632911392	13.1019954935204\\
4.64556962025316	14.1485414753089\\
5.1506329113924	15.0321749381771\\
5.65569620253164	15.7787702136137\\
6.16075949367089	16.4111699063552\\
6.66582278481013	16.9488012517641\\
7.17088607594937	17.4078480209758\\
7.67594936708861	17.8016407279623\\
8.18101265822785	18.1410988425467\\
8.68607594936709	18.4351496860706\\
9.19113924050633	18.6910950577611\\
9.69620253164557	18.9149190692658\\
10.2012658227848	19.111540650716\\
10.7063291139241	19.2850179364562\\
11.2113924050633	19.4387125049298\\
11.7164556962025	19.5754209080006\\
12.2215189873418	19.6974799103209\\
12.726582278481	19.8068507610729\\
13.2316455696203	19.9051868076004\\
13.7367088607595	19.9938878925129\\
14.2417721518987	20.0741442609476\\
14.746835443038	20.1469721292391\\
15.2518987341772	20.2132426093745\\
15.7569620253165	20.2737053237941\\
16.2620253164557	20.3290077629578\\
16.7670886075949	20.3797112173018\\
17.2721518987342	20.4263039424663\\
17.7772151898734	20.4692120813982\\
18.2822784810127	20.508808760814\\
18.7873417721519	20.5454216960621\\
19.2924050632911	20.5793395726268\\
19.7974683544304	20.6108174204733\\
20.3025316455696	20.6400811561294\\
20.8075949367089	20.6673314345189\\
21.3126582278481	20.6927469262781\\
21.8177215189873	20.7164871152152\\
22.3227848101266	20.7386946936133\\
22.8278481012658	20.7594976193777\\
23.3329113924051	20.7790108879305\\
23.8379746835443	20.7973380627238\\
24.3430379746835	20.8145726008738\\
24.8481012658228	20.8307990043881\\
25.353164556962	20.8460938224969\\
25.8582278481013	20.8605265265154\\
26.3632911392405	20.874160275288\\
26.8683544303797	20.8870525864585\\
27.373417721519	20.8992559264857\\
27.8784810126582	20.910818230374\\
28.3835443037975	20.9217833604641\\
28.8886075949367	20.9321915122634\\
29.3936708860759	20.9420795741463\\
29.8987341772152	20.9514814467845\\
30.4037974683544	20.9604283273516\\
30.9088607594937	20.9689489628452\\
31.4139240506329	20.9770698762853\\
31.9189873417722	20.9848155690392\\
32.4240506329114	20.992208702096\\
32.9291139240506	20.9992702587456\\
33.4341772151899	21.0060196907999\\
33.9392405063291	21.0124750502235\\
34.4443037974684	21.0186531078076\\
34.9493670886076	21.0245694603176\\
35.4544303797468	21.0302386273722\\
35.9594936708861	21.0356741391577\\
36.4645569620253	21.0408886159521\\
36.9696202531646	21.0458938403181\\
37.4746835443038	21.0507008227246\\
37.979746835443	21.0553198612687\\
38.4848101265823	21.0597605960957\\
38.9898734177215	21.0640320590444\\
39.4949367088608	21.0681427189919\\
40	21.072100523314\\
};
\addlegendentry{Local Upper Bound via \eqref{eq:LMMSE}}

\end{axis}
\end{tikzpicture}%
  \caption{ Comparing proposed bounds for a prior distribution uniform on $K$-ball versus the radius $R$.   }
  \label{fig:uniform}
\end{figure}
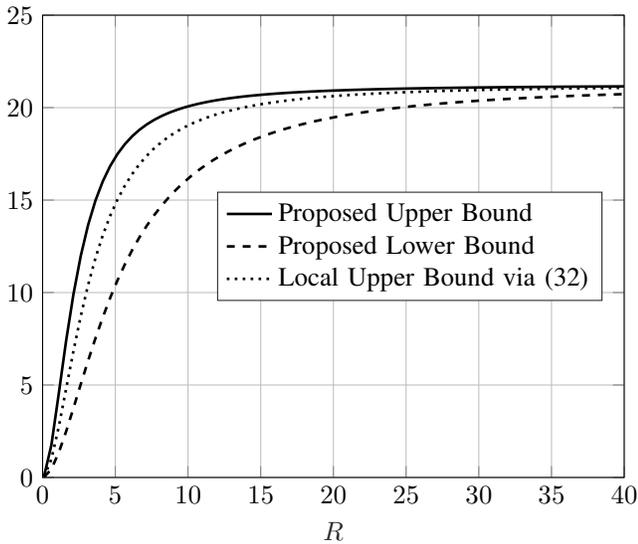

Another important feature of the proposed bounds is that they hold for prior distributions that do not necessarily have a well-defined Fisher information. Note that, as a consequence of Stem's inequality, the finiteness of Fisher information implies finite  $D_{\text{KL}}(P_X\|Q)$.   The converse statement, however, is not true, and there are distributions without well-defined Fisher information but with finete $D_{\text{KL}}(P_X\|Q)$. The practical implications of this phenomenon is that, our bounds hold for a larger set of distributions. This property of  the proposed bounds has been discussed in detail in \cite{FaussDytso2018_TSP}.  An example of such a prior distribution is a uniform distribution over a $K$-ball.   We demonstrate this in Fig.~\ref{fig:uniform}, where the plots of our lower and upper bounds versus the radius of the $K$-ball using  parameters in \eqref{eq:channelParameters} is shown. It is interesting to observe that the larger the radius of the $K$-ball the better is the performance of the proposed lower and upper bounds.

\section{Concluding remarks}
\label{sec:conclusion}

Upper and lower bounds on the weighted sum of MMSEs for additive Gaussian noise channels have been derived. It has been shown that these bounds take the coupling between the individual MMSEs into account and thereby are significantly tighter than the existing bounds. Examples have been provided to show how the presented results can be particularly useful for the design and analysis of sensor networks for distributed estimation, where the weighted sum of MMSEs can be used to monitor the performance, establish operating regions and to design robust local estimators. Further insights into the robustness of the local estimators for distributed setups require a more careful analysis and will be subject of future research.

\section*{Acknowledgment}
The work of A. Dytso and H. V. Poor was supported by the U. S. National Science Foundation under Grant CCF–1513915.  The work of Nagananda K. G. was supported by the European Research Council under grant agreement 715111.

\bibliographystyle{IEEEbib}
\bibliography{refs}

\end{document}